\documentclass[amsmath,amssymb,twocolumn]{revtex4}
\usepackage[parfill]{parskip}    
\usepackage{graphicx}
\usepackage{amssymb}
\usepackage{epstopdf}
\usepackage{color}
\usepackage{graphicx}
\usepackage{pdfpages}

\begin{document}

\title{Gravitational decoupling in $2+1$ dimensional space--times with cosmological term}
\author{Ernesto Contreras {\footnote{On leave 
from Universidad Central de Venezuela}
\footnote{econtreras@yachaytech.edu.ec}} }
\address{Yachay Tech University, School of Physical Sciences \& Nanotechnology, 100119-Urcuqu\'i, Ecuador\\}
\begin{abstract}
In this work we implement the Minimal Geometric Deformation method to obtain the
isotropic sector and the decoupler matter content of any anisotropic solution of the 
Einstein field equations with cosmological constant in $2+1$ dimensional space--times. We obtain that the solutions of both sectors can be expressed analytically in terms of the metric functions of the original anisotropic solutions instead of formal integral as in its $3+1$ counterpart. As a particular example we study a regular black hole solution and we show 
that, depending on the sign of the cosmological constant, the solutions correspond
to regular black holes violating the null energy condition or to a non--regular black hole without exotic hair. The exotic/non--exotic and the regular/non--regular black hole dualities are discussed.

\end{abstract}
\maketitle

\section{Introduction}\label{intro}
The interest in the Minimal Geometric Deformation (MGD)
\cite{randall1999a,randall1999b,antoniadis1990,antoniadis1998,ovalle2008,ovalle2009,ovalle2010,casadio2012,ovalle2013,ovalle2013a,
casadio2014,casadio2015,ovalle2015,casadio2015b,
ovalle2016, cavalcanti2016,casadio2016a,ovalle2017,
rocha2017a,rocha2017b,casadio2017a,ovalle2018,estrada2018,ovalle2018a,lasheras2018,gabbanelli2018,
sharif2018,fernandez2018,fernandez2018b,contreras2018,estrada,contreras2018a,morales}
as a powerful method to decouple the Einstein field equations to obtain new solutions 
\cite{ovalle2017,ovalle2018,estrada2018,ovalle2018a,lasheras2018,gabbanelli2018,
sharif2018,contreras2018,contreras2018a,rincon2018,ovalleplb,contreras2018c,tello2018,tello2018a,
tello2018b} has considerably increased. Among the main applications of the method we find
studies of local anisotropies in spherically
symmetric systems \cite{lasheras2018,gabbanelli2018,tello2018,tello2018a,
tello2018b}, hairy black holes \cite{ovalle2018a} and new anisotropic  solutions
in $2+1$ dimensional space--times \cite{contreras2018,rincon2018}.

The method has been extended to solve the inverse problem \cite{contreras2018a}, namely, given any anisotropic solution of the Einstein field equations it is possible to recover the isotropic 
sector and the decoupler matter content which, after gravitational interaction, led to the anisotropic configuration. In that work, it was found that, for an anisotropic solution
with exotic matter sector (negative energy density), the free parameters involved in the MGD can be setted such that both the isotropic and the decoupler sectors satisfy all the energy conditions. It was the first time that a kind of exotic/non--exotic matter was found using the method. 

As another extension of MGD and the inverse problem, in Ref. \cite{contreras2018c} 
the method have been studied considering Einstein's equations with cosmological constant and 
implemented in a polytropic black hole which is a solution with a matter content
satisfying all the energy conditions. The main finding there was that
the isotropic sector is deeply linked with the appearance of
exotic matter, although it can be located inside the horizon.
In this sense, this work shows again how the apparition of exotic matter seems unavoidable but one could, in principle, control the energy conditions by tuning
the isotropy/anisotropy of a black hole solution.

The MGD-decoupling have been implemented in $2+1$ circularly symmetric
and static space–times obtaining that both the isotropic
and the anisotropic sector fulfil Einstein field equations in
contrast to the cases studied in $3+1$ dimensions, where
the anisotropic sector satisfies “quasi-Einstein” field
equations. In this sense, the isotropic and the decoupler sector leads to
a pair of new solution of
Einstein’s equations, one for each source. 

In this work we study MGD in $2+1$ circularly symmetric and static space–times with different purposes and interests.
First, as a difference to the previous work (see Ref. \cite{contreras2018}), we consider $2+1$ Einstein's equation with cosmological constant. This is because, 
as the BTZ is a vacuum solution of this configuration, the set of equations coming from MGD  method could serve as the starting point to extend interior $2+1$ solutions to anisotropic domains taking into account suitable matching conditions between the compact objects and a BTZ vacuum. Second, we study the inverse MGD problem to explore, among other aspects, the exotic/non--exotic matter content duality previously reported in the $3+1$ dimensional case \cite{contreras2018a,contreras2018c}. As we shall see later, the inverse method leads to more tractable expressions to deal with because they correspond to exact analytic instead to formal equations as previously reported \cite{contreras2018a,contreras2018c}.

This work is organized as follows. In the next section we briefly review the MGD-decoupling method.
Then, in section \ref{iso}, we obtain the isotropic sector and the decoupler matter content
considering a regular black hole as anisotropic solution. In section \ref{ec} we study the energy conditions to explore the apparition of exotic mater in the solutions and some final comments and conclusion are in the last section.

\section{Einstein Equations in $2+1$ space--time dimensions}\label{mgd}
In a previous work we considered the MGD-method with cosmological constant \cite{contreras2018c}. In this work study the 
$2+1$ dimensional case. 

Let us consider the Einstein field equations
\begin{eqnarray}\label{einsorig}
R_{\mu\nu}-\frac{1}{2}R g_{\mu\nu}+\Lambda g_{\mu\nu}=\kappa^{2}T_{\mu\nu}^{tot},
\end{eqnarray}
and assume that the total energy-momentum tensor is given by
\begin{eqnarray}\label{total}
T_{\mu\nu}^{(tot)}=T_{\mu\nu}^{(m)}+\theta_{\mu\nu},
\end{eqnarray}
As usual, the energy--momentum tensor for a perfect fluid is given by
$T^{\mu(m)}_{\nu}=diag(-\rho,p,p)$ and the decoupler matter content reads 
$\theta^{\mu}_{\nu}=diag(-\rho^{\theta},p_{r}^{\theta},p_{\perp}^{\theta})$. 
In what follows, we shall assume circularly symmetric space--times with a line element parametrized as
\begin{eqnarray}\label{le}
ds^{2}=-e^{\nu}dt^{2}+e^{\lambda}dr^{2}+r^{2}d\phi^{2},
\end{eqnarray}
where $\nu$ and $\lambda$ are functions of the radial coordinate $r$ only. 
Considering Eq. (\ref{le}) as a solution of the Einstein Field Equations, we obtain
\begin{eqnarray}
\kappa ^2 \tilde{\rho}&=&-\Lambda+\frac{e^{-\lambda} \lambda'}{2 r}\label{eins1}\\
\kappa ^2 \tilde{p}_{r}&=&\Lambda+\frac{e^{-\lambda} \nu '}{2 r}\label{eins2}\\
\kappa ^2 \tilde{p}_{\perp}&=&\Lambda+\frac{1}{4} e^{-\lambda} \left(-\lambda ' \nu '+2 \nu ''+\nu '^2\right)\label{eins3}
\end{eqnarray}
where the prime denotes derivation with respect to the radial coordinate and we have defined
\begin{eqnarray}
\tilde{\rho}&=&\rho+\rho^{\theta}\label{rot}\\
\tilde{p}_{r}&=&p+p_{r}^{\theta}\label{prt}\\
\tilde{p}_{\perp}&=&p+p_{\perp}^{\theta}.\label{ppt}
\end{eqnarray} 

The next step consists in decoupling the Einstein Field
Equations (\ref{eins1}), (\ref{eins2}) and (\ref{eins3}) by implementing the minimal deformation
\begin{eqnarray}\label{def}
e^{-\lambda}&=&\mu +\alpha f.
\end{eqnarray}
As usual, Eq. {\ref{def}} leads to
two sets of differential equations: one describing an isotropic system sourced by
the conserved energy--momentum tensor of a perfect fluid $T^{\mu(m)}_{\nu}$ an the other
set corresponding to Einstein field equations sourced by $\theta_{\mu\nu}$. 
Now, as in a previous work \cite{contreras2018c}, we interpret the cosmological constant 
as an isotropic fluid, so we include the $\Lambda$--term in the isotropic sector and
we obtain
\begin{eqnarray}
\kappa ^2\rho &=& \frac{-2 \Lambda  r-\mu '}{2 r}\label{iso1}\\
\kappa ^2 p&=& \frac{2 \Lambda  r+\mu \nu '}{2  r}\label{iso2}\\
\kappa ^2 p&=&\frac{4 \Lambda +\mu ' \nu '+\mu \left(2 \nu ''+\nu '^2\right)}{4 }
,\label{iso3}
\end{eqnarray}
for the perfect fluid and
\begin{eqnarray}
\kappa ^2\rho^{\theta}&=&-\frac{f'}{2 r}\label{aniso1}\\
\kappa^{2} p_{r}^{\theta}&=&\frac{f \nu '}{2 r}\label{aniso2}\\
\kappa^{2} p_{\perp}^{\theta}&=&\frac{f' \nu '+f \left(2 \nu ''+\nu '^2\right)}{4},\label{aniso3}
\end{eqnarray}
for the anisotropic system \footnote{In what follows we shall assume $\kappa^{2}=8\pi$ }. Note that that the addition of the cosmological constant only affects 
the isotropic sector because  Eqs. (\ref{aniso1}), (\ref{aniso2})
and (\ref{aniso3}) remain unchanged. 

For the inverse problem we shall apply the same strategy followed in reference \cite{contreras2018a}, namely, we implement the constraint
\begin{eqnarray}\label{cons}
\tilde{p}_{\perp}-\tilde{p}_{r}=\alpha(p^{\theta}_{\perp}-p^{\theta}_{r})
\end{eqnarray}
It is remarkable that, unlike the $3+1$ case, the solution for $f(r)$ obtained from the constraint 
(\ref{cons}) is an exact expression in terms of $\nu,\lambda$, instead of a combination of formal integrals as found in \cite{contreras2018a}. In fact,
\begin{eqnarray}\label{f}
f=\frac{c_1 r^2 e^{-\nu }}{\nu '^2}+\frac{e^{-\lambda}}{\alpha }
\end{eqnarray}
where $c_1$ is a constant of integration. Now, from Eq. (\ref{def}), we obtain 
\begin{eqnarray}\label{mu}
\mu=-\frac{\alpha  c_1 r^2 e^{-\nu}}{\nu '^2}
\end{eqnarray}
Note that somehow, the constant $c_1$ controls the geometric deformation: if $c_1\to0$ there is not
deformation at all. 

The matter content of the isotropic sector reads
\begin{eqnarray}
\rho&=&-\Lambda-\frac{\alpha  c_1 e^{-\nu} \left(2 r \nu ''+\nu ' \left(r \nu '-2\right)\right)}{2 \nu '^3}\\
p&=&\Lambda -\frac{\alpha  c_1 r e^{-\nu}}{2 \nu '},
\end{eqnarray} 
and  the decoupler matter content satisfies
\begin{eqnarray}
\rho^{\theta}&=&\frac{1}{2} \left(\frac{c_1 e^{-\nu} \left(2 r \nu ''+\nu ' \left(r \nu '-2\right)\right)}{\nu '^3}+\frac{e^{-\lambda} \lambda '}{\alpha  r}\right)\\
p_{r}^{\theta}&=&\frac{c_1 r e^{-\nu}}{2 \nu '}+\frac{e^{-\lambda} \nu '}{2 \alpha r}\\
p_{\perp}^{\theta}&=&\frac{1}{4} \left(\frac{2 c_1 r e^{-\nu}}{\nu '}+\frac{e^{-\lambda} \left(-\lambda '\nu '+2 \nu ''+\nu '^2\right)}{\alpha }\right)
\end{eqnarray}

At this point, some comments are in order. First,
Eqs. (\ref{iso1}), (\ref{iso2}) and (\ref{iso3}) correspond to Einstein field equations
with cosmological constant for a perfect fluid. Second, Eqs. (\ref{aniso1}), (\ref{aniso2}) and (\ref{aniso3}) corresponds to 
Einstein field equations without cosmological constant and anisotropic decoupler fluid. Note that
this set of equations together with those of the isotropic sector allows us to decouple Einstein's equation with cosmological constant in $2+1$ dimensional space--times for any anisotropic fluid. What is more, 
the above expressions can be used to extend isotropic solutions embedded in a BTZ vacuum to anisotropic domains after the implementation of suitable matching conditions. Finally, that
the inverse problem leads to exact analytical expressions entails the ``isotropization'' of a
broader set of systems than its $3+1$ dimensional counterpart which is given in terms of
formal integrals. In this sense, the inverse problem in $2+1$ dimensions can be implemented starting
from any anisotropic solutions at difference to the $3+1$ case where depending on the particular
form of the anisotropic solution, the inverse problem would yield to formal instead exact analytical solutions. In the next section we shall implement the inverse problem using a well known
regular and circularly symmetric black hole solution as anisotropic system.

\section{Isotropic sector of a regular black hole in 2+1 dimensions}\label{iso}
In this section we shall illustrate the inverse MGD problem using as anisotropic solution a
regular black hole metric. The reason to consider a regular black hole solution is twofold: to explore the conditions for the apparition of exotic matter and to study if the MGD inverse problem
can affect the regularity of the solution.

Let us consider the regular black hole solution \cite{garcia2017} with metric functions
\begin{eqnarray}
e^\nu&=& -M-\Lambda  r^2-q^{2}\log\left(a^2+r^2\right)\\
e^{-\lambda}&=&-M-\Lambda  r^2-q^{2}\log \left(a^2+r^2\right)
\end{eqnarray}
where $M$, $\Lambda$, $a$ and $q$ are free parameters. This geometry is sustained by a matter
content given by
\begin{eqnarray}
\tilde{\rho}&=&\frac{q^2}{8\pi\left(a^2+r^2\right)}\\
\tilde{p}_{r}&=&-\frac{q^2}{8\pi\left(a^2+r^2\right)}\\
\tilde{p}_{\perp}&=&\frac{q^2 \left(r^2-a^2\right)}{8\pi\left(a^2+r^2\right)^2}
\end{eqnarray}
This solution corresponds to a black hole whenever $ -M-\Lambda  r^2-q^{2}\log\left(a^2+r^2\right)=0$ leads to two real roots (or one real root in the extreme case) for
some values of the parameters $\{M,\Lambda,q,a\}$ \cite{garcia2017}. What is more, the solution
is regular everywhere, which can be deduced from the invariants
\begin{eqnarray}
R&=&\frac{2 q^2 \left(3 a^2+r^2\right)}{\left(a^2+r^2\right)^2}+6 \Lambda\\
Ricc=\mathcal{K}&=&\frac{4 q^4 \left(3 a^4+2 a^2 r^2+r^4\right)}{\left(a^2+r^2\right)^4}\nonumber\\
&&+\frac{8 \Lambda  q^2 \left(3 a^2+r^2\right)}{\left(a^2+r^2\right)^2}
+12 \Lambda ^2,
\end{eqnarray}
where $R$, $Ricc$ and $\mathcal{K}$ correspond to the Ricci, Ricci squared and  Kretschmann scalar respectively.

From now on we shall apply the inverse MGD problem to obtain the isotropic generator and the 
decoupler matter content associated with this regular black hole solution.
From Eq. (\ref{f}), the decoupler function $f$ reads
\begin{eqnarray}
f&=&-\frac{c_1 \left(a^2+r^2\right)^2 \left(q^2 \log \left(a^2+r^2\right)+M+\Lambda  r^2\right)}{4 \left(\Lambda  \left(a^2+r^2\right)+q^2\right)^2}\nonumber\\
&&-\frac{q^2 \log \left(a^2+r^2\right)+M+\Lambda  r^2}{\alpha }
\end{eqnarray}
Now, from Eq. (\ref{mu}), the radial metric function of the isotropic sector is given by
\begin{eqnarray}
\mu=\frac{\alpha  c_1 \left(a^2+r^2\right)^2 \left(q^2 \log \left(a^2+r^2\right)+M+\Lambda  r^2\right)}{4 \left(\Lambda  \left(a^2+r^2\right)+q^2\right)^2}
\end{eqnarray}
Replacing the above result in the set of isotropic Einstein equations, Eqs. (\ref{iso1}), (\ref{iso2}) and (\ref{iso3}), the perfect fluid reads
\begin{eqnarray}
&&\rho=
-\frac{\alpha  c_1 a^{2}_{r}\left(2 \Lambda  q^2 \left(a^{2}_{r}+r^2\right)+\Lambda ^2 a_{r}^2+2 M q^2+q^4\right)}{32\pi \lambda_{r}^3}\nonumber\\
&&-\frac{2 \alpha  c_1 q^4 a^{2}_{r} \log a^{2}_{r}+4 \Lambda  \lambda_{r}^3}{
32\pi\lambda_{r}^3}\\
&&p=\frac{\alpha  c_1 \left(a^2+r^2\right)}{32\pi \left(\Lambda  \left(a^2+r^2\right)+q^2\right)}+\frac{\Lambda }{8\pi}
\end{eqnarray}
where
\begin{eqnarray}
a_{r}^{2}&:=&a^{2}+r^2\\
\Lambda_{r}&:=&\Lambda  a^{2}_{r}+q^2
\end{eqnarray}
At this point some comments are in order. First, note that, as in the $3+1$ case, the inverse problem
does not modify the position  of the killing horizon. In fact, the horizon appears whenever
$-M-\Lambda  r^2-q^{2}\log\left(a^2+r^2\right)=0$ which, as discussed above, leads to one or two 
real roots for the black hole solution. Second, the regularity of the solution depends on the positivity of the parameter $\Lambda$. 
In fact, the invariants
\begin{eqnarray}
R&=&-\alpha  c_{1}a_{r}^{2}\left(\frac{  \left(2 M q^2+3 q^4\right)}{2 \Lambda_{r}^3}
-\frac{2 q^4 \log a_{r}^{2}}{2 \Lambda_{r}^3}\right.\nonumber\\
&&\left.-\frac{2 \Lambda  q^2 \left(3 a^2+4 r^2\right)}{2 \Lambda_{r}^3}-\frac{3 \Lambda ^2 a_{r}^{3}}{2 \Lambda_{r}^3}\right)\\
Ricc=\mathcal{K}&=&\frac{\alpha ^2 c_1^2 a_{r}^4 \left(\mathcal{F}+2 \Lambda _r^4\right)}{4 \Lambda _r^6}
\end{eqnarray}
where 
\begin{eqnarray}
\mathcal{F}=\left(2 q^4 \log a_{r}^{2}+2 \Lambda  q^2 a_{r}^{2}+
\Lambda ^2 a_{r}^4+2 M q^2+q^4\right)^2,
\end{eqnarray}
reveal that the solution is regular everywhere whenever 
$\Lambda_{r}=\Lambda  \left(a^2+r^2\right)+q^2\ne 0$, which can be satisfied if
$\Lambda>0$. In particular, from the horizon condition we can
obtain the condition for $\Lambda>0$ in terms of the other parameters as
\begin{eqnarray}
\Lambda=-\frac{M+q^{2}\log(a^{2}+r_{H}^{2})}{r_{H}^{2}}
\end{eqnarray}
from where it must be imposed that $r_{H}^{2}+a^{2}<1$ and $q^{2}|\log(a^{2}+r_{H}^{2})|>M$, with 
$r_{H}$ the horizon radius. In this case, the solution corresponds to a regular isotropic black hole solution. 


Note that, in the case $\Lambda<0$
the solution has a critical radius $r_{c}$ as frequently found in the application of MGD. This 
result would lead to a naked singularity for $r_{c}>r_{H}$ or to a non regular black hole solution for $r_{c}<r_{H}$. In the last case, we say that the isotropic sector of the regular black hole corresponds to a non-regular black hole solution.

Now we focus our attention into the decoupler sector. In this case, the metric functions  are 
$\{\nu,f\}$ and the decoupler matter content reads
\begin{eqnarray}
\rho^{\theta}&=&\frac{c_1 a_r^2 \left(2 \Lambda  q^2 \left(a_{r}^2+r^{2}\right)+\Lambda ^2 a_r^4+2 M q^2+q^4\right)}{4 \Lambda _r^3}\nonumber\\
&&+\frac{q^2}{\alpha  a_r^2}
+\frac{c_1 q^4 a_r^2 \log a_r^2}{2 \Lambda _r^3}+\frac{\Lambda }{\alpha }\\
p_{r}^{\theta}&=&-\frac{c_1 a_r^2}{4 \Lambda _r}-\frac{q^2}{\alpha  a_r^2}+\frac{\Lambda }{\alpha }\\
p_{\perp}^{\theta}&=&\frac{q^2 \left(r^2-a^2\right)}{\alpha  a_r^4}-\frac{c_1 a_r^2}{4 \Lambda _r}-\frac{\Lambda }{\alpha}
\end{eqnarray}

The above solution corresponds to an anisotropic regular black hole solution for $\Lambda>0$. In fact, the solution
has a killing horizon when $-M-\Lambda  r^2-q^{2}\log\left(a^2+r^2\right)=0$. The invariants are given by
\begin{eqnarray}
R&=&\frac{\mathcal{H}_{1}}{\left(a^2+r^2\right)^2 \left(\Lambda  \left(a^2+r^2\right)+q^2\right)^3}\\
Ricc&=&\frac{\mathcal{H}_{2}}{\left(a^2+r^2\right)^4 \left(\Lambda  \left(a^2+r^2\right)+q^2\right)^6}\\
\mathcal{K}&=&\frac{\mathcal{H}_{3}}{\left(a^2+r^2\right)^4 \left(\Lambda  \left(a^2+r^2\right)+q^2\right)^6}
\end{eqnarray}
where $\mathcal{H}_{1}$, $\mathcal{H}_{2}$ and $\mathcal{H}_{3}$ are (too long) regular functions
in terms of polynomials of $r$ and $\log(a^{2}+r^{2})$. Note that, as discussed above, the 
regularity of the solution depends on the sign of $\Lambda$. More precisely, for $\Lambda<0$
the solution is the one of a non--regular black hole solution.

It is worth mentioning that the solutions obtained here could be considered as ``hairy'' black holes
and that the nature of such a hair fields depends on the energy conditions that we shall discuss in
what follows.

\section{Energy conditions}\label{ec}
In this section we study the energy conditions of the obtained solution for the distinct cases 
outlined before.
\subsection{Case I. $\Lambda>0$}
\begin{figure}[h!]
\centering
\includegraphics[width=\linewidth]{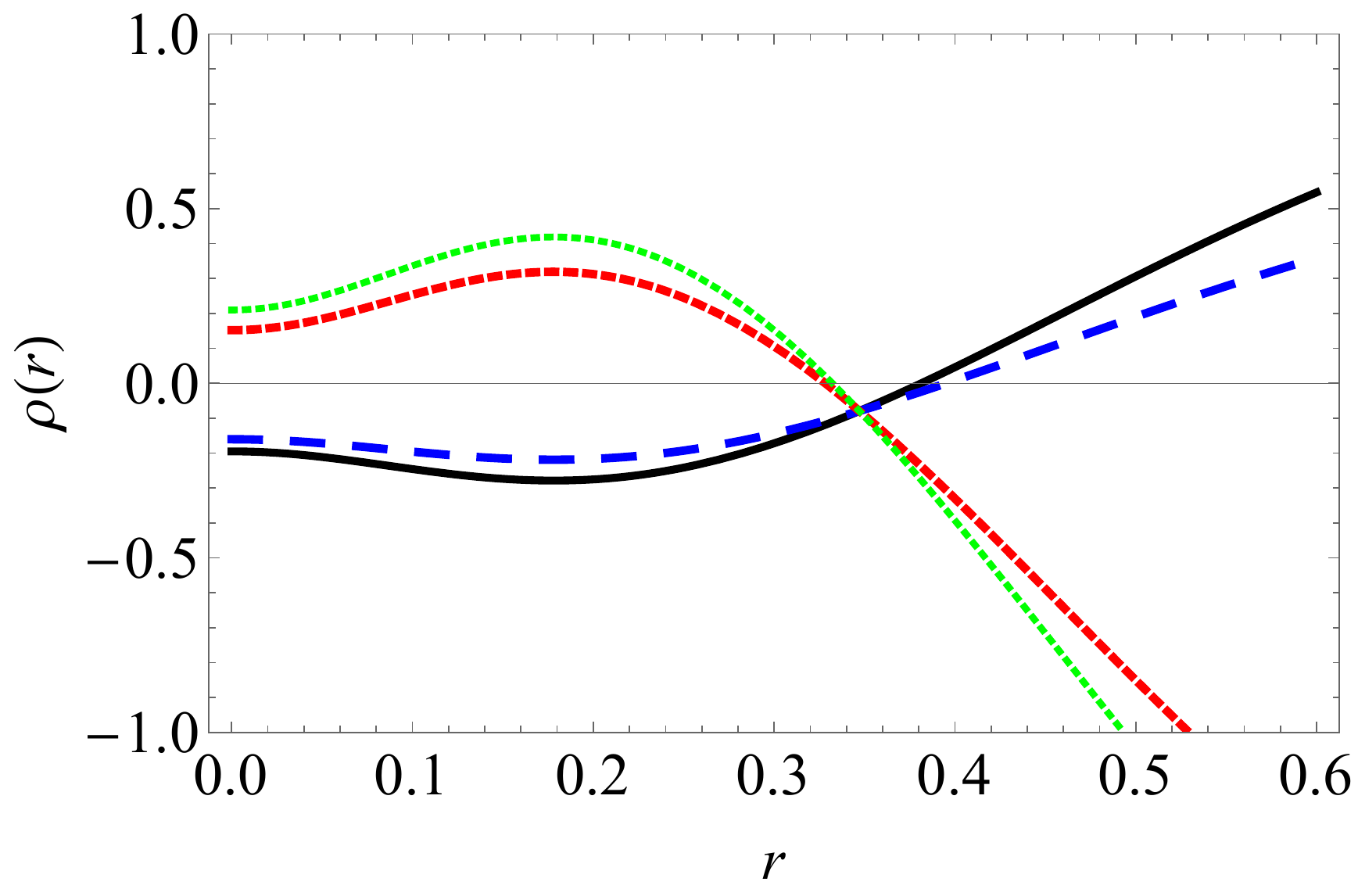}
\caption{\label{fig1} 
Energy density for for $\alpha c_{1}=-200$ (black solid line), $\alpha c_{1}=-140$ (dashed blue line)
$\alpha c_{1}=400$ (short dashed red line) and $\alpha c_{1}=500$ (dotted green line).
}
\end{figure}

As previously discussed, this case corresponds to regular black hole solutions for the isotropic and decoupler sector with the horizon radius located at $-M-\Lambda r_{H}^{2}-q^{2}\log(a^{2}+r_{H}^{2})=0$. Given the nature of the solution, a numerical analysis is mandatory. However, setting suitable values of $M,\Lambda,q,a$ to obtain real horizons and $\Lambda>0$, we infer that the behaviour of the energy density can be written as
\begin{eqnarray}
&&\lim\limits_{r\to0}\rho=A\alpha  c_1-B\\
&&\lim\limits_{r\to\infty}\rho=-C \alpha  c_1-B
\end{eqnarray}
with $A$, $B$, $C$ real and positive numbers. In particular, for
$\Lambda=2$, $M=1$, $q=1$ and $a= 0.1$ we obtain
\begin{eqnarray}
&&\lim\limits_{r\to0}\rho=0.000578336 \alpha  c_1-0.0795775\\
&&\lim\limits_{r\to\infty}\rho=-0.00497359 \alpha  c_1-0.0795775
\end{eqnarray} 
Note that the apparition of exotic matter is unavoidable. In fact, if $\alpha  c_1$ is a positive
(negative) quantity such that
$\lim\limits_{r\to0}\rho>0\ (\lim\limits_{r\to0}\rho<0)$, it is obtained that 
$\lim\limits_{r\to\infty}\rho>0\ (\lim\limits_{r\to\infty}\rho<0)$ necessarily. In figure \ref{fig1}
we show the profile of the energy density for some values of $\alpha c_{1}$.

\begin{figure}[h!]
\centering
\includegraphics[width=\linewidth]{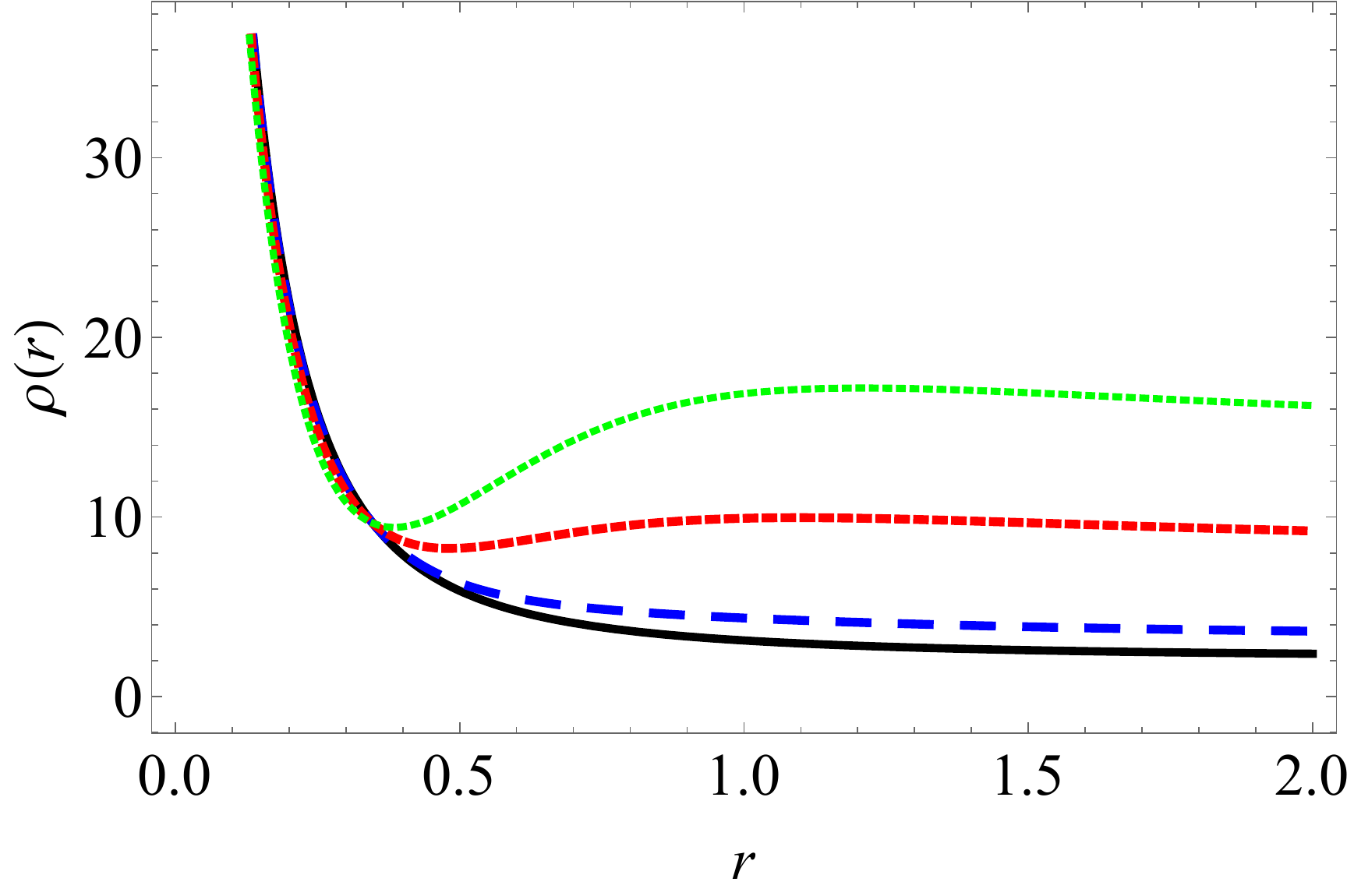}
\caption{\label{fig2} 
Energy density for for $c_{1}=1$ (black solid line), $c_{1}=10$ (dashed blue line)
$c_{1}=50$ (short dashed red line) and $c_{1}=100$ (dotted green line).
}
\end{figure}

For the decoupler sector we obtain that the negative values for the energy density can be avoided with a suitable choice of the parameter. Without loss of generality, let us set
$\Lambda=2$, $M=1$, $q=1$ and $a= 0.1$ to ensure that the horizon radius is real and positive. Now
\begin{eqnarray}
\lim\limits_{r\to0}\rho^{\theta}&=&\frac{102.}{\alpha }-0.0145352 c_1\\
\lim\limits_{r\to\infty}\rho^{\theta}&=&\frac{2}{\alpha }+\frac{c_1}{8}
\end{eqnarray}
Note that for $\alpha$ and $c_{1}$ positive values, the exotic matter content can be avoided whenever
$c_{1}\le\frac{7017.46}{\alpha }$. In figure \ref{fig2} we show the profile of 
$\rho^{\theta}$ for $\alpha=1$ and some values for $c_{1}$.

We would like to conclude this section by emphasizing that for $\Lambda>0$ the exotic
matter can be partially avoided. In fact, while the perfect fluid solution contain negative energy 
density for suitable values of $M,q,\Lambda,a$, the apparition of exotic matter in the decoupler sector can be circumvented for certain values of the parameters involved. In the next section we shall study the energy conditions for $\Lambda<0$.

\subsection{Case II. $\Lambda<0$}
\begin{figure}[h!]
\centering
\includegraphics[width=\linewidth]{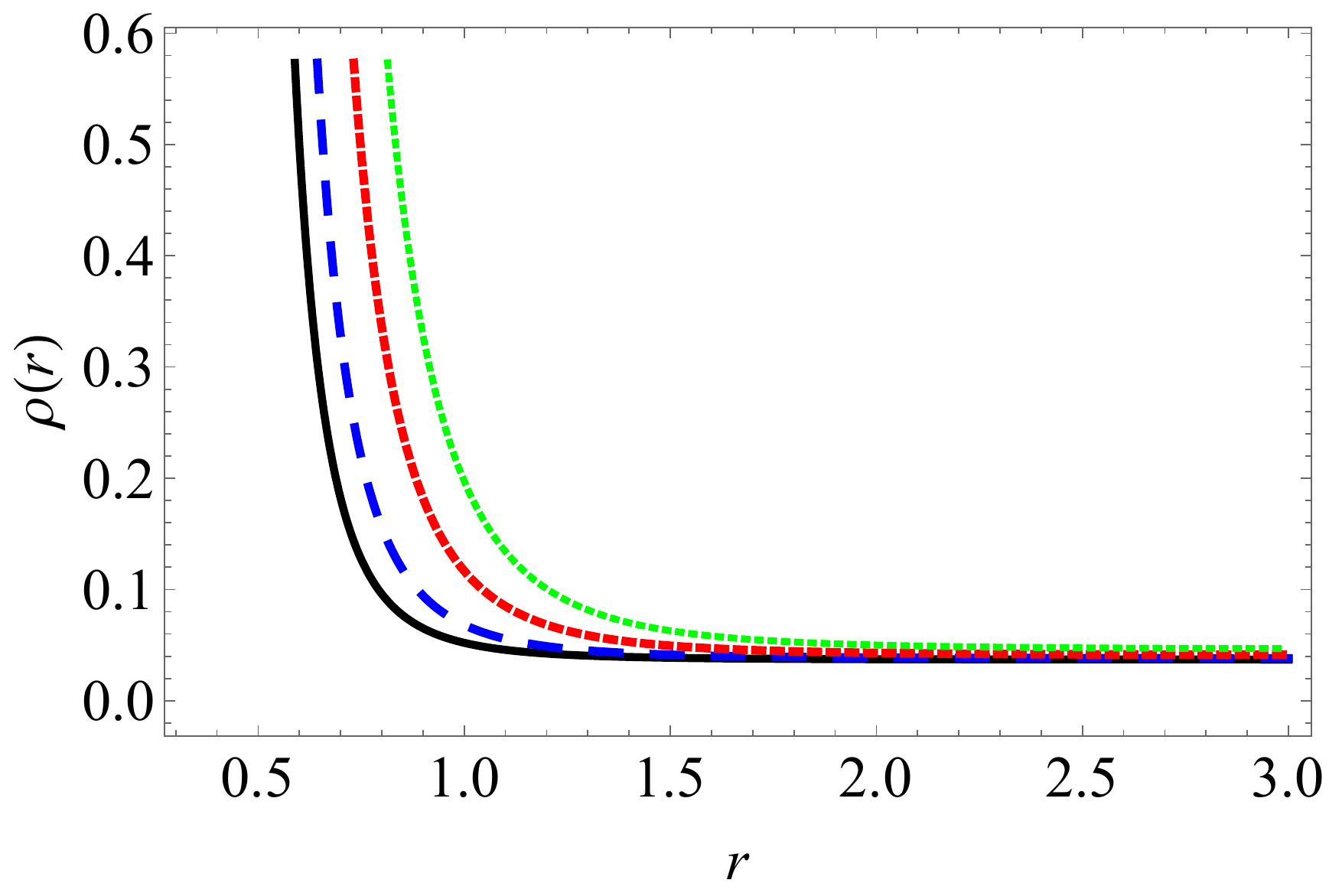}
\caption{\label{fig3} 
Energy density for for $\alpha c_{1}=0.1$ (black solid line), $\alpha c_{1}=0.2$ (dashed blue line)
$\alpha c_{1}=0.5$ (short dashed red line) and $\alpha c_{1}=1$ (dotted green line).
}
\end{figure}

In this case, $\Lambda<0$ as long as $r_{H}^{2}+a^{2}<1$ and\break
$q^{2}|\log(a^{2}+r_{H}^{2})|>M$
or $r^{2}+a^{2}>1$. For example, we can choose the values $M=2$, $q=1$, $a=1$ from where
$\Lambda=-0.902359$. For these values, the horizon radius is located at $r_{H}=2$ and the critical radius is at
$r_{c}=0.328946$ such that the solution corresponds to a non--regular black hole solution. 
For this particular choosing of the parameters and for $\alpha c_{1}>0$, the asymptotic behaviour of the energy density is given by
\begin{eqnarray}
&&\lim\limits_{r\to r_{c}}\rho=\infty\\
&&\lim\limits_{r\to\infty}\rho=0.0110235 \alpha  c_1+0.0359037.
\end{eqnarray}
In figure in figure \ref{fig3} we show the profile of $\rho$ as a function of the radial coordinate for different values of $\alpha c_{1}>0$.

\begin{figure}
\centering
\includegraphics[width=\linewidth]{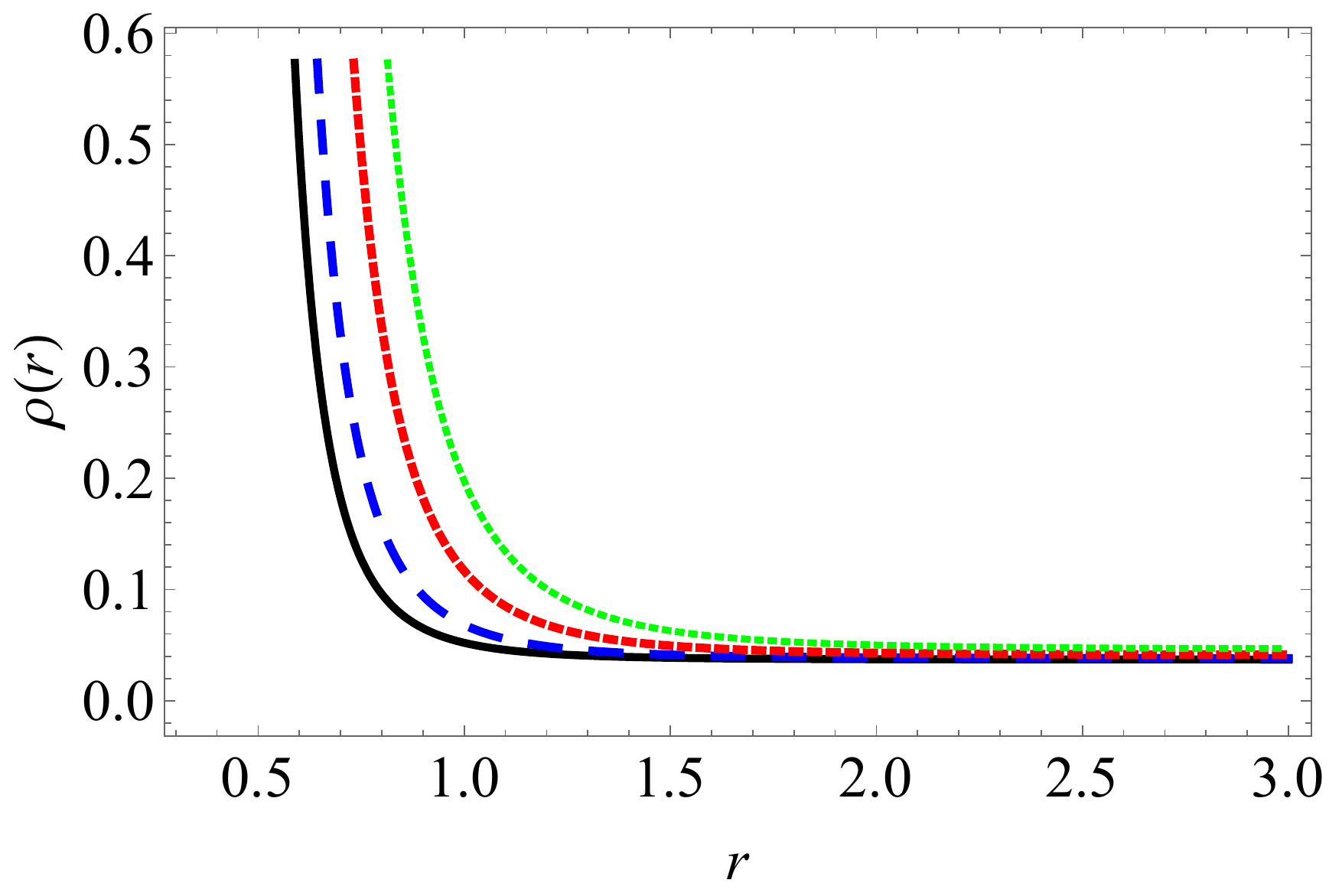}
\caption{\label{fig4} 
Energy density for for $c_{1}=-1$ (black solid line), $c_{1}=-2$ (dashed blue line)
$c_{1}=-3$ (short dashed red line) and $c_{1}=-4$ (dotted green line).
}
\end{figure}

Now, let us turn out our attention in the decoupler sector. For
$M=2$, $q=1$, $a=1$ and $\Lambda=-0.902359$ we obtain
\begin{eqnarray}
\lim\limits_{r\to\infty}\rho^{\theta}=-\frac{0.902359}{\alpha }-0.277051 c_1
\end{eqnarray}
so that for $\alpha<0$, $c_{1}<0$ the energy density reach a positive value asymptotically . In particular for $\alpha=-1$ we obtain that
\begin{eqnarray}
\lim\limits_{r\to r_{c}}\rho^{\theta}\to\infty
\end{eqnarray}
and we obtain that the exotic matter can be avoided. In figure \ref{fig4} we show the profile of the energy density $\rho^{\theta}$ for $\alpha=-1$ and different values of $c_{1}$.

At his point a couple of comments are in order. First, note that in both cases (isotropic and decoupler sector) the exotic content
can be avoided. Second, for suitable choices of the parameters, the solution corresponds to a 
non--regular black hole containing a non--vanishing critical radius. In this sense, we conclude that
although the exotic/non--exotic duality can be circumvented for $\Lambda<0$, the nature of 
the black hole solution of the isotropic and decoupler sector leads to a kind of regular/non--regular duality.

\section{Conclusions}\label{remarks}
In this work we have extended the Minimal Geometric Deformation method in $2+1$ dimensional space--times to decouple the Einstein field equations including cosmological constant. We obtained that
the isotropic sector obeys Einstein's equation with cosmological constant but the decoupler 
part consists in a system without cosmological term. In this sense, we can combine any $2+1$ 
isotropic, static and circularly symmetric interior solution of the Einstein field equations with cosmological constant embedded in a BTZ vacuum with certain decoupler matter solution and suitable matching conditions to obtain new anisotropic interior solutions in the three dimensional realm.

We showed that the inverse problem leads to exact analytical solutions for the decoupling and the isotropic metric in terms of the original anisotropic solution instead to formal integrals obtained in the $3+1$ counterpart. The scope of this result to obtain analytical solutions is broad. Indeed, it can be implemented taking into account any anisotropic solution as the starting point because
it does not involve formal integrals as the $3+1$ case.  As a particular example we implemented the inverse problem to a regular $2+1$ black hole solution.
We obtain that for a positive cosmological constant the isotropic sector corresponds to
a regular isotropic black hole in presence of a ``exotic'' hair (negative energy density), and the decoupler sector is a regular anisotropic black hole which, under certain circumstances, can be surrounded by a matter content with positive energy density so that the apparition of exotic matter can be avoided. For negative cosmological constant both the isotropic and the decoupler sector
corresponds to non--regular black hole solution where the existence of exotic hair can be avoided
with a suitable choice of the free parameters. It is worth mentioning that on one hand the non--regular black hole solution is singular in a non--vanishing radius as often occur in the implementation of the Minimal Geometric Deformation protocol. On the other hand, the exotic matter can be avoided but the price that it has to be paid is that the the solutions are not regular 
anymore. In this sense, the kind of exotic/non-exotic matter duality appearing in previous works 
transmute to a regular/non-regular duality in the cases where the exotic content can be avoided.
\section{Acknowledgement}
The author would like to acknowledge Pedro Bargue\~no for fruitful discussions.

\end{document}